\documentclass[paper]{JHEP3}
\JHEPspecialurl{http://jhep.sissa.it/JOURNAL/JHEP3.tar.gz}
\usepackage{epsfig,multicol}
\usepackage{cite} 
\usepackage{amsmath,amssymb}  
\usepackage{bm} 
\usepackage{enumerate}   
\usepackage{type1cm}
   
\newcommand\fverb{\setbox\pippobox=\hbox\bgroup\verb}
\newcommand\fverbdo{\egroup\medskip\noindent%
			\fbox{\unhbox\pippobox}\ }
\newcommand\fverbit{\egroup\item[\fbox{\unhbox\pippobox}]}
\newcommand {\beq}{\begin{equation}}
\newcommand {\eeq}{\end{equation}}
\newcommand {\beqa}{\begin{eqnarray}}
\newcommand {\eeqa}{\end{eqnarray}}

\newcommand {\tr}{{\rm tr\,}}

\newcommand {\ee}{\mbox{e}}

\def\laplace{{\kern1pt\vbox{\hrule height 1.2pt\hbox{\vrule width 1.2pt\hskip
  3pt\vbox{\vskip 6pt}\hskip 3pt\vrule width 0.6pt}\hrule height 0.6pt}
  \kern1pt}}
\def\scriptlap{{\kern1pt\vbox{\hrule height 0.8pt\hbox{\vrule width 0.8pt
  \hskip2pt\vbox{\vskip 4pt}\hskip 2pt\vrule width 0.4pt}\hrule height 0.4pt}
  \kern1pt}}

\def\roughly#1{\raise.3ex\hbox{$#1$\kern-.75em\lower1ex\hbox{$\sim$}}}

\preprint{KEK-TH-1160}
\title{
Phase structure
of matrix quantum mechanics
at finite temperature
}
\author{ 
Naoyuki Kawahara${}^{ab}$\,,
Jun Nishimura${}^{ac}$ 
and Shingo Takeuchi${}^{c}$ 
\vspace*{0.5cm} \\  
\llap{$^a$}High Energy Accelerator Research Organization (KEK),\\
Tsukuba, Ibaraki, 305-0801, Japan  \\
\llap{$^b$}Department of Physics, Kyushu University,
Fukuoka 812-8581, Japan \\ 
\llap{$^c$}Department of Particle and Nuclear Physics,\\
Graduate University for Advanced Studies (SOKENDAI),\\
Tsukuba, Ibaraki, 305-0801, Japan
\vspace*{0.5cm} \\ 
\email{kawahara@post.kek.jp,
jnishi@post.kek.jp, shingo@post.kek.jp}
}

\abstract{
We study matrix quantum mechanics
at finite temperature by Monte Carlo simulation.
The model 
is obtained by dimensionally reducing 10d 
U($N$) pure Yang-Mills theory to 1d.
Following Aharony et al., one can view
the same model as describing
the high temperature regime of (1+1)d 
U($N$) {\em super} Yang-Mills theory on a circle.
In this interpretation
an analog of the deconfinement transition was conjectured
to be a continuation of the black-hole/black-string
transition in the dual gravity theory.
Our detailed analysis in the critical regime up to $N=32$
suggests the existence of the non-uniform phase,
in which the eigenvalue distribution 
of the holonomy matrix
is non-uniform but gapless.
The transition to the gapped phase is of second order.
The internal energy is constant
(giving the ground state energy)
in the uniform phase, and rises quadratically
in the non-uniform phase, which implies that the
transition between these two phases is of third order.
}

\begin{document} 

\section{Introduction} 
\label{section:Introduction}

%%%

Recently large-$N$ gauge theories are playing 
increasingly
%more and more 
important roles in string theory.
% from two different points of view.
%One is that they 
One of the crucial discoveries was that
U($N$) gauge theory appears
as a low energy effective theory
\cite{Witten:1995im}
for a stack of $N$ D-branes \cite{Polchinski:1995mt}
in string theory.
This led to various interesting 
conjectures concerning large-$N$ gauge theories.
For instance, it is conjectured that
large-$N$ gauge theories obtained
by dimensionally reducing 10d U($N$) super Yang-Mills
theory to $D=0,1,2$ dimensions
provide non-perturbative formulations of 
superstring/M theories.
These are called Matrix theory ($D=1$) \cite{BFSS}, 
the IIB matrix model ($D=0$) \cite{IKKT} and 
matrix string theory ($D=2$) \cite{DVV}, respectively.
Another type of conjectures
%, which is studied extensively during the decade,
asserts the duality between strongly coupled 
large-$N$ gauge theory and
%have dual descriptions in terms of 
weakly coupled supergravity.
In the AdS/CFT correspondence
%\footnote{Recently a certain
%version of the equivalence 
%has been derived from the symmetry of the worldsheet
%theory under anisotropic scale transformations \cite{Kawai:2007ek}.
%}
\cite{Maldacena:1997re,Gubser:1998bc,Witten:1998qj},
for instance,
it is conjectured that
4-dimensional ${\rm U}(N)$ $\mathcal{N}=4$ 
super Yang-Mills theory is dual to 
the type IIB supergravity on ${\rm AdS}_5 \times {\rm S}^5$.
%% This duality can be used in both ways.
%% One may hope to study 
%% the former
%% %strongly coupled large-$N$ gauge theory
%% using the latter,
%% %weakly coupled supergravity,
%% but one may also try to understand
%% blackhole physics, for instance, by considering
%% %in terms of 
%% gauge theory at finite temperature \cite{Witten:1998zw}.
%% %% Large-$N$ gauge theories
%% %% are therefore expected to provide a natural
%% %% quantum description of gravity phenomena
%% %% such as blackhole evaporation and big bang,
Generalizing this correspondence to
the finite temperature case, it was argued that
the deconfinement phase transition on the gauge theory side
corresponds to the Hawking-Page transition on the gravity side
\cite{Witten:1998zw}. 
This has been further extended to non-conformal gauge theories
including supersymmetric matrix quantum mechanics 
\cite{Itzhaki:1998dd}.
See refs.\ \cite{Barbon:1998cr,Martinec,KLL,Harmark:2002tr,%
Aharony:2004ig,Aharony,%
Aharony:2005bm,Aharony4,wadia,Harmark:2007md}
for extensive studies on the relationship between
large-$N$ gauge theories at finite temperature
and the black-hole physics.
%% It is generally believed that
%% large-$N$ gauge theories at finite temperature
%% are related to black-hole physics. (See, for instance,
%% ref.\ \cite{Barbon:1998cr,Martinec,KLL,Aharony:2004ig,Aharony,%
%% Aharony4,wadia,Harmark:2007md}).

In this paper we study
matrix quantum mechanics
%``bosonic Matrix theory'' 
at finite temperature
by Monte Carlo simulation.
The model 
%takes the form of matrix quantum mechanics 
is obtained formally by dimensionally reducing 10d U($N$) 
pure Yang-Mills theory to 1d,
and it may be viewed\footnote{One may 
also consider the model as
the ``bosonic Matrix theory'' \cite{latticeBFSS},
but in that case, the compactified Euclidean time 
direction actually corresponds to a light-cone coordinate,
%should be interpreted as one of the light-cone 
%directions, 
which might make the meaning of 
``finite temperature'' a bit subtle
unlike in the other interpretations. 
See refs.\ \cite{BFKS},
%,KS,Horowitz:1997fr,Das:1997tk,AMS},
however.
}
as the bosonic part of 
the low energy effective theory of D-particles 
in type IIA superstring theory \cite{Witten:1995im}.
%but one may also view it as
%omitting the fermionic part.
%
%Previous Monte Carlo studies take the 
%
%% In either of these interpretations, including
%% the fermionic matrices would be 
%% an important future direction.
%However, viewing the temperature axis as a spatial axis,
By considering the Euclidean time direction
as a spatial direction instead,
one can view the bosonic model as describing
the high temperature regime of 
(1+1)d U($N$) ${\cal N}=8$ {\em super} Yang-Mills theory 
on a circle.
In this interpretation 
an analog of the deconfinement transition was
speculated  \cite{Aharony:2004ig}
to be a continuation
of the black-hole/black-string phase transition
in the dual gravity theory\footnote{Some 
of the results in refs.\ 
\cite{Aharony:2004ig}
were already anticipated in a pioneering work \cite{Harmark:2002tr}
on the phase structure of black holes on a circle.
There, the relationship of the supergravity solutions to 
non- and near-extremal branes on a circle and 
to the corresponding dual non-gravitational theories
are discussed.
}.
%we can discuss the relationship
%to the black hole-black string phase transition
%in the dual gravity theory \cite{Aharony:2004ig}.

%% Compared with the previous works 
%% \cite{latticeBFSS,Aharony:2004ig},
%% we take the lattice spacing
%% in the time direction to be much smaller
%% so that 
%% our Monte Carlo results represent
%% % practically
%% %essentially
%% the continuum limit.
%% This is confirmed by
%% % the fact the
%% precise agreement of the asymptotic behavior 
%% at high temperature
%% with the high temperature expansion
%% up to the next-leading order.
%% The Eguchi-Kawai equivalence observed
%% at low temperature
%% % to very good accuracy.
%% %This 
%% enables precise determination of the ground state
%% of the quantum mechanical system.

Unlike previous works,
our detailed analysis in the critical regime 
up to $N=32$
%The eigenvalue distribution of the holonomy matrix
%can be fitted well to
%the Gross-Witten form \cite{GWW},
%, suggesting that
%we are observing the large-$N$ behavior.
%In particular, 
%and we observe
% the second order phase transition
%between 
suggests the existence of the non-uniform phase,
in which the eigenvalue distribution of the 
holonomy matrix is non-uniform but gapless.
The transition to the gapped phase appears to be
of second order.
At low temperature,
the internal energy is constant (giving the ground state energy)
as a result of the Eguchi-Kawai equivalence.
We use this property to identify the uniform phase.
As one enters the non-uniform phase increasing
the temperature, the internal
energy starts to rise quadratically.
This implies that the transition between the uniform phase
and the non-uniform phase is of third order.
These observations select one of the two scenarios
suggested by the Landau-Ginzburg analysis \cite{Aharony:2004ig}.

The rest of this paper is organized as follows.
In section \ref{section:model} we define
the model and the observables we study.
% and discuss its basic property.
In section \ref{section:phase} we present 
an overview of the phase diagram as seen from
explicit results for the observables.
In section \ref{section:critical} we focus on the
critical regime and study the phase structure in more
detail.
In section \ref{section:GL} we use
our results to speculate on the 
phase diagram of 
%review the interpretation
%of the model as the high temperature limit of 
(1+1)d super Yang-Mills theory on a circle.
%, and discuss the implication of our results. 
Section \ref{Summary} is devoted to a summary and
discussions.
%In section \ref{sec:boundary} we study...
In appendix 
\ref{section:EO} we derive a formula
for the internal energy, which is used for its
evaluation.
In appendix \ref{section:algorithm}
we present the details of
%explain the algorithm we use for 
our simulation.

%\section{Matrix quantum mechanics at finite temperature}
\section{The model and observables}
\label{section:model}

The model we study in this paper is 
defined by the action
\beq 
\label{action}
S = \frac{1}{g^2} \int_0^\beta \!\!dt \, \tr
\left\{ \frac{1}{2} 
\Bigl( D_t X_i(t) \Bigr) ^2 -\frac{1}{4}
%\Bigl( 
[X_i(t),X_j(t)]^2 
%\Bigr) 
\right\} \ ,
\eeq
where $D_t$ represents the covariant derivative
$D_t \equiv 
\partial_t - i \, [A(t), \ \cdot \ ] $.
%% \\
%% \label{partition}
%% Z[\beta] &=& \int[dX][dA]\exp(-S[\beta])\,,
%% \eeqa
The one-dimensional fields $A(t)$ and
$X_i(t)$ $(i=1,2,\cdots,9)$ 
are $N \times N$ Hermitian matrices, which
can be regarded as the gauge field and 
nine adjoint scalars, respectively. 
The model can be formally obtained 
by dimensionally reducing 
10d U($N$) pure Yang-Mills theory to 1 dimension.
%%Since the trace part of the gauge field decouples,
%we restrict $A(t)$ to be traceless.
%
%As a result, the gauge symmetry of the action we consider 
%becomes SU($N$). 
The Euclidean time $t$ has a finite extent $\beta$, 
which corresponds to the inverse temperature
$\beta = 1/T$,
and all the fields obey periodic boundary conditions.

The 't Hooft coupling constant $\lambda \equiv g^2 N$ 
has the dimension of $(\rm{mass})^3$,  
and we fix it when we take the large-$N$ limit.\footnote{This 
is different from the
decompactifying limit \cite{BFSS} in the Matrix theory.}
%Since the 't Hooft coupling constant has the
%dimension of $(\rm{mass})^3$,  
The physical properties of the system depend only on
the dimensionless effective coupling constant
given by\footnote{One can confirm this statement by rescaling 
the fields and the coordinate $t$ appropriately
so that all the $\lambda$ and $T$ dependence
appears in the combination of eq.\ (\ref{lam-eff}).}
%$\lambda_{\rm eff} = \frac{\lambda}{T^3}$.
\beq
\lambda_{\rm eff} \equiv \frac{\lambda}{T^3} \ .
\label{lam-eff}
\eeq
%
%% \beq
%% g_{\rm eff}^2=\frac{\lambda}{T^3} \ .
%% \eeq
In what follows we set
% the 't Hooft coupling constant to 
$\lambda=1$
without loss of generality.
%which implies that we measure the temperature
%in units of $\lambda^{1/3}$.
%
%See Appendix for
%the details of the Monte Carlo simulation performed to
%study the system (\ref{action}).

It is known that the bosonic matrix quantum mechanics
undergoes a phase transition \cite{latticeBFSS,Aharony:2004ig,%
Kawahara:2007nw} at some critical temperature,
which can be interpreted as the Hagedorn transition
in string theory \cite{Atick:1988si,Aharony}.
This transition is associated with the spontaneous breakdown 
of the U(1) symmetry 
\beq
A(t) \mapsto A(t) + \alpha {\bf 1} \ ,
\label{U1-sym}
\eeq
and therefore it is analogous to
the deconfinement transition in ordinary gauge theories.
The order parameter is given by the Polyakov line
\beqa
\label{pol-def}
P &\equiv& \frac{1}{N} {\rm tr} \, U \ , \\
U &\equiv& 
%\left|
%{\rm tr} \, 
\mathcal{P} \exp\left(i \int_0^{\beta} \!\!dt A(t) \right)
%\right| 
\ ,
\label{pol-def2}
\eeqa
%% \beq
%% P \equiv \frac{1}{N} 
%% %\left|
%% {\rm tr} \, 
%% \mathcal{P} \exp\left(i \int_0^{\beta} \!\!dt A(t) \right)
%% %\right| 
%% \ ,
%% \label{pol-def}
%% \eeq
where the symbol $\mathcal{P}\exp$ represents the path-ordered
exponential and the unitary matrix $U$ is called
the holonomy matrix.
%As is always the case with the spontaneous breaking of symmetries,
%we will simply obtain $\langle P \rangle \equiv 0$ 
%Note that $0 \le \langle |P| \rangle \le 1$. 
%
In section \ref{section:critical}
we present numerical results suggesting that
the ``deconfined phase'' is further divided into
two phases.

As a fundamental quantity in thermodynamics,
the free energy 
${\cal F} \equiv - \frac{1}{\beta} \ln Z(\beta)$
is defined in terms of the partition function
%given in the present model as
\beq
Z(\beta) = \int [{\cal D} X]_\beta 
[{\cal D} A]_\beta \, \ee^{- S(\beta)} \ ,
\label{def-part-fn}
\eeq
where the suffix of the measure $[ \ \cdot  \ ]_\beta $
represents the period of the field
to be path-integrated.
However, the free energy ${\cal F}$ cannot be calculated 
straightforwardly
by Monte Carlo simulation because that would
require evaluation of the partition function 
$Z(\beta)$.
%instead of some expectation values.
We therefore study the internal energy defined by
\beq
E \equiv \frac{d}{d \beta} (\beta {\cal F}) 
  = - \frac{d}{d \beta} \log Z(\beta) \ ,
\label{defE}
\eeq
which has 
%the same almost 
equivalent 
information as 
the free energy (given that ${\cal F}=E$ at $T=0$).
Note also that the internal energy at $T=0$
provides the ground state energy of the quantum
mechanical system.
In appendix \ref{section:EO} we derive a formula
\beqa
\label{E=dbF}
\frac{1}{N^2} E &=& \frac{3}{4} \, \langle F^2 \rangle  \ , \\
F^2 &\equiv &
-  \frac{1}{N}
\int_0^\beta  \!\!dt  \, \tr
\Bigl( [X_i  (t  ),X_j (t )]^2 \Bigr) \ ,
\label{F2def}
\eeqa
where the symbol
$\langle \ \cdot \ \rangle$ 
represents the expectation value with respect
to $Z(\beta)$.
%Eq.\ (\ref{E=dbF})
This formula
enables us to calculate the internal energy $E$
directly by Monte Carlo simulation. 
As yet another quantity, we study
\beq
\label{def_R2}
R^2 \equiv
\frac{1}{N\beta} \int_0^{\beta} \!\!dt \, \tr (X_i)^2 \ ,
\eeq
which represents the ``extent of space''.

The details of our simulation 
are given in appendix \ref{section:algorithm}.
% \ref{app:algorithm}.
The number of sites $N_{t}$ in the Euclidean time direction
is taken to be the smallest integer that satisfies
$N_{t} \ge \frac{1}{0.02 T}$
% otherwise.
for figs.\ \ref{Fig_PL} and \ref{fig:BHTE} ,
and $N_{t} \ge \frac{1}{0.05 T}$ for figs.\
\ref{fig:PL2}, \ref{fig:BHTE2} and \ref{fig:EVD}.
%in the critical regime,
%and 
%ranges from 17 to 125 in the region of temperature 
%presented in this paper.
%
%This implies that the lattice spacing $a=\frac{1}{N_{t} T}$
%is less than $0.02$,
%which is small enough to extract the continuum limit
%as we know from the study of an analogous model 
%\cite{Kawahara:2005an}.
%
%our previous work
% of an analogous model 
This corresponds to taking the lattice spacing
to be $a \simeq 0.02$ and $a \simeq 0.05$, respectively. 
See ref.\ \cite{Kawahara:2005an} for a systematic
study of the finite lattice-spacing effects
in
%It is also known from the study of 
an analogous model.
%\cite{Kawahara:2005an}
%that results for the lattice spacing $a \simeq 0.05$ 
%
%we used for $N=32$ 
%is sufficiently close to the continuum limit.
%small to see the continuum limit.
%as we know 
%\cite{Kawahara:2005an}.
% than the previous works.
Note that the previous work \cite{Aharony:2004ig}
used $N_{t} = 5$ at any temperature.
%(In the critical regime, our $N_{t}$ is 4 times larger
%than the previous work .)

\section{An overview of the phase diagram}
\label{section:phase}

\FIGURE{
\epsfig{file=
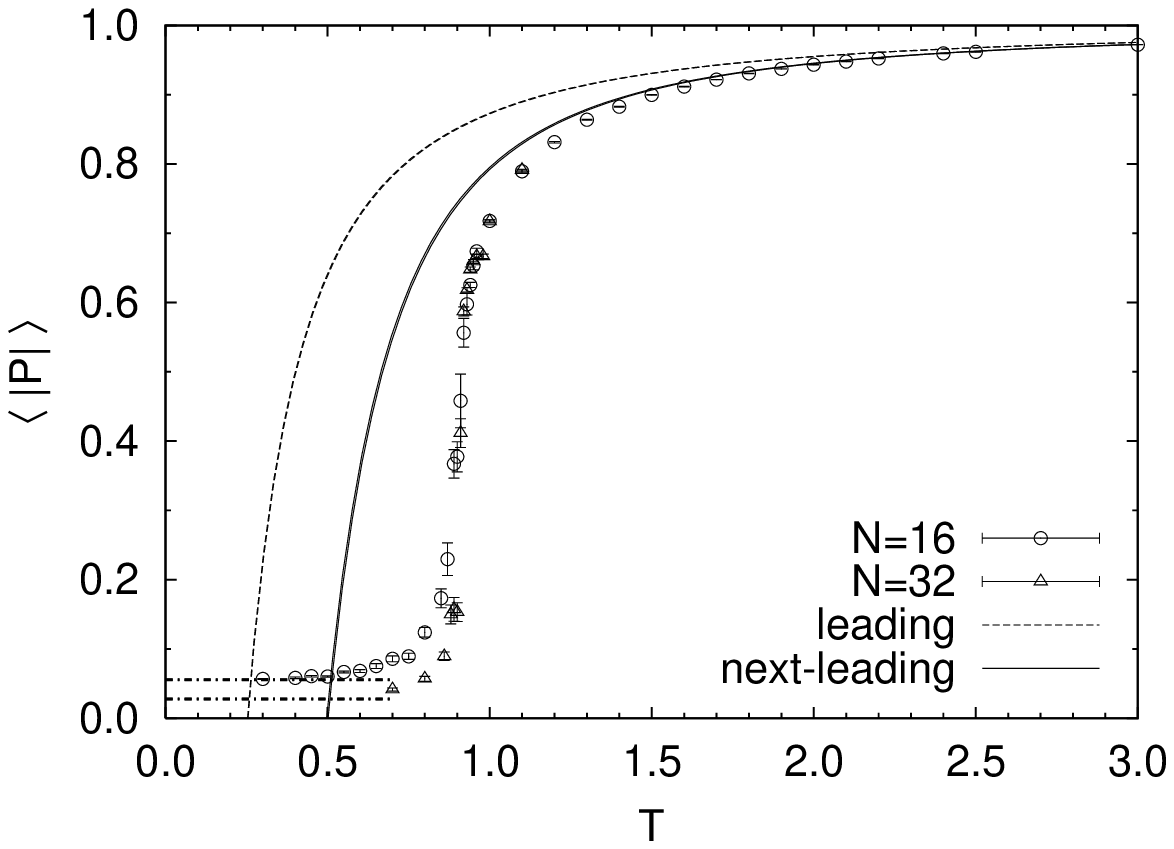, width=.49\textwidth
%7.4cm
}
\caption{
The Polyakov line $\langle |P| \rangle$ is plotted 
against $T$ for $N=16$, and in
the critical regime, also for $N=32$.
%We also plot $N=32$ data points in
%significant finite $N$ effects are expected.
The dashed line and the solid line
represent the results of
the high temperature expansion \cite{highT}
for $N=16$,
which are obtained at the leading order
and at the next-leading order, respectively.}
\label{Fig_PL}
}  

In this section we present an overview of
the phase diagram of the model (\ref{action}).
Fig.\ \ref{Fig_PL} shows the results for the Polyakov 
line.
It
%The order parameter $\langle |P| \rangle$ 
changes drastically at $T \simeq 0.9$.
For $T \gtrsim 0.9$, the results for $N=16$ and $N=32$
lie on top of each other,  
%hence showing convergence at large $N$.
and approaches the maximum value 1 as $T$ increases.
The results at $T \gtrsim 2$ are nicely reproduced
by the high temperature expansion\footnote{The 
results of the high temperature expansion \cite{highT}
are plotted with the double lines
showing the statistical errors 
coming from the Monte Carlo integration over the
zero modes. 
In some cases, the errors are so small that one cannot
recognize the lines as two separate ones.
%% Actually there is no statistical
%% error in the leading order result 
%% for the internal energy because
%% in that case the integration over the zero modes
%% can be performed exactly.
} including the next-leading order terms \cite{highT}.
For $T \lesssim 0.9$,
the expectation value $\langle |P| \rangle$
%the Polyakov line 
takes small values, which
seem to vanish in the large-$N$ limit as 
$1/N$. 

In fact, the expectation value 
$\langle |P| \rangle$ in the $T \rightarrow 0$ limit
seems to agree well with
%can be reproduced by 
the finite-$N$ results obtained
by generating the holonomy matrix $U$ randomly 
using the Haar measure, which are represented
by the horizontal dash-dotted lines in
fig.\ \ref{Fig_PL} for $N=16,32$.
The latter results are seen to be
proportional to $1/N$ for $N=8,16,32,64$.
In this way, we can understand the observed
$1/N$ behavior of $\langle |P| \rangle$
in the original model at low temperature.
On the other hand, finite $N$ effects 
at $T \gtrsim 0.9$ seems to be much smaller.
This can be understood from the fact that
in this regime the phenomenological description
of the holonomy matrix is given by the Gross-Witten
model (\ref{GWW-def}),
which has finite $N$ effects of
the order of O($1/N^2$) for $N \gg \sqrt{\kappa}$.
%% The phenomenological description also explains why
%% the Polyakov line $\langle |P| \rangle$
%% has small finite $N$ effects in the 
%% iantells Gross-Witten model also explains 

%This $1/N$ behavior can be understood intuitively.
%Up to $N=32$, we do not see any evidence for discontinuity.

%Suppose the eigenvalues of the holonomy matrix $U$
%in eq.\ (\ref{pol-def}) are randomly distributed on a unit circle.
%Then one obtains $\langle |P| \rangle \sim {\rm O}(\frac{1}{N})$.

%% \DOUBLEFIGURE[bht]
%% {E.eps, width=.49\textwidth}
%% {R2.eps, width=.49\textwidth}
%% {Similar plot as fig.\ \ref{Fig_PL}
%% for the internal energy $E$.
%% \label{fig:BHTE}}
%% {Similar plot as fig.\ \ref{Fig_PL}
%% for the ``extent of space'' $\langle R^2 \rangle$.
%% \label{Fig_R2}} 

  \FIGURE{
\epsfig{file=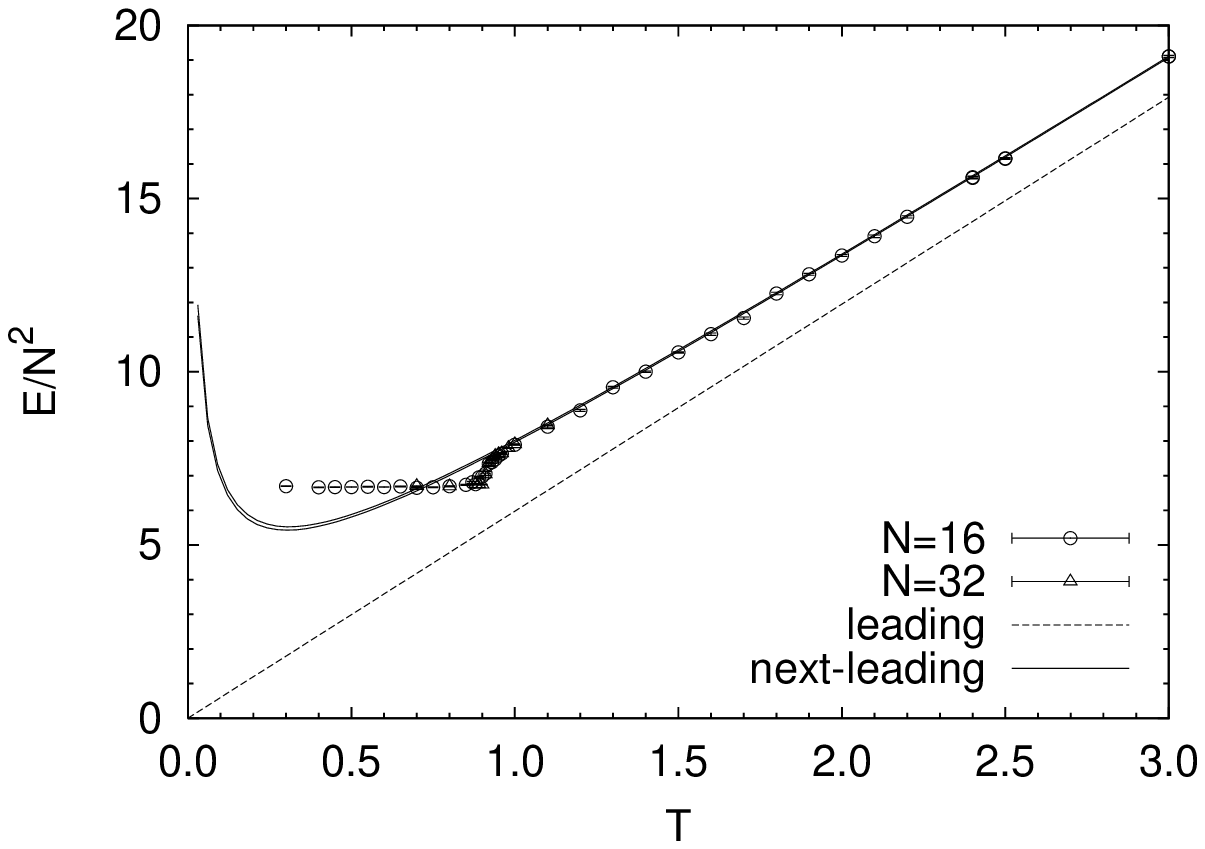,width=7.4cm}
\epsfig{file=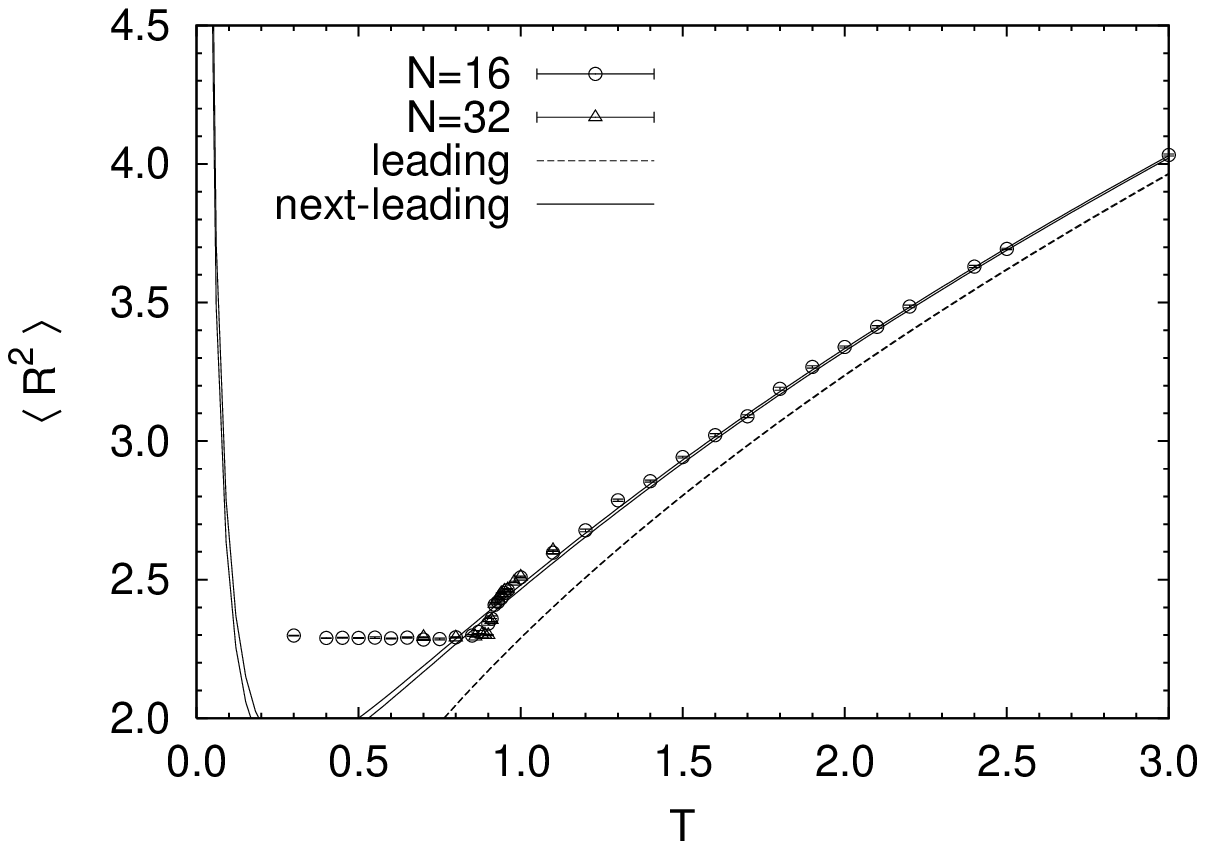,width=7.4cm}
   \caption{
The same as fig.\ \ref{Fig_PL} but
for the internal energy $\frac{1}{N^2} E$ (left)
and for the ``extent of space'' $\langle R^2 \rangle$
(right).
}
  \label{fig:BHTE}
}
In fig.\ \ref{fig:BHTE} we
plot the
internal energy obtained from the formula (\ref{E=dbF}),
and 
%in fig.\ \ref{Fig_R2} we plot 
the ``extent of space'' $\langle R^2 \rangle$.
For $T \lesssim 0.9$,
we observe that
the results
%the internal energy $E$ and the extent of space
%$\langle R^2 \rangle$ 
are independent of $T$.
This can be understood as a consequence of
the Eguchi-Kawai equivalence.
In general it states 
that the expectation value of
%volume independence of 
a single-trace
operator in $D$-dimensional
U($N$) gauge theory 
is independent of the volume in the large-$N$ limit,
provided that the ${\rm U}(1)^D$ symmetry is 
%$({\rm Z}_N)^D$ symmetry is 
not spontaneously broken\footnote{In 3d 
and 4d pure SU($N$) gauge theory on a torus,
the SSB occurs at finite 
volume \cite{Narayanan:2003fc,Kiskis:2003rd},
and the relation to the deconfinement transition
at finite temperature \cite{Lucini:2002ku} is discussed.
See also refs.\ \cite{Cohen:2004cd,Furuuchi:2005qm,%
Polchinski:1991tw} for related works.
} \cite{EK}.
%In the present case, 
The $T$-independence of $\lim_{N \rightarrow \infty}
\frac{1}{N^2} E$ 
%at low $T$
at low temperature
%the Eguchi-Kawai equivalence 
%observed at low temperature 
%
%has a physical interpretation.
%
%in the confined phase
%
%It is a consequence of the fact that
can also be viewed as a consequence of the fact that
%has a physical meaning that
only U($N$) invariant
%color-singlet 
states appear in the low energy spectrum
in a confining theory.

\FIGURE{
\epsfig{file=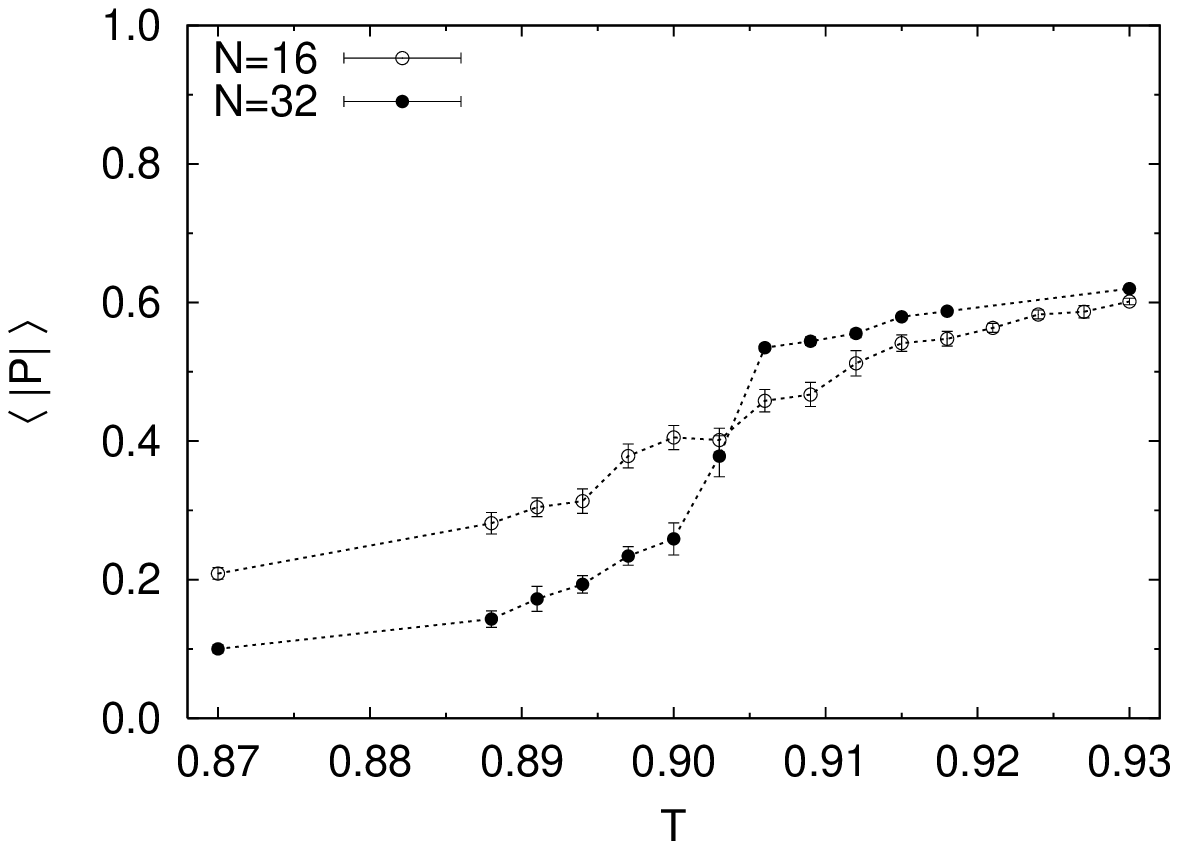, width=.49\textwidth
%7.4cm
}
\caption{
The Polyakov line $\langle |P| \rangle$ is plotted 
against $T$ in the critical regime for $N=16$ and $N=32$.
The dotted lines are drawn to guide the eye.
}
\label{fig:PL2}
}

\section{Closer look at the critical regime}
%\section{Detailed analysis in the critical regime}
\label{section:critical}

%%%
In this section 
we focus on the critical regime
and study the phase structure in more detail.
%we present more detailed
%analysis in the critical regime.

In fig.\ \ref{fig:PL2} we plot the
Polyakov line against temperature
magnifying the critical regime.
For $N=16$ the Polyakov line
changes smoothly with $T$.
However, for $N=32$
% we observe that 
the behavior of the Polyakov line
changes drastically at 
$T \sim T_{{\rm c}1}\equiv 0.905(2)$.
In fig.\ \ref{fig:BHTE2}
we plot the internal energy $E$
and the ``extent of space'' $\langle R^2 \rangle$
in the critical regime.
The qualitative behavior is similar to the
Polyakov line.
%For $N=16$ the results changes smoothly with $T$,
%but for $N=32$ we observe a rapid change of the behavior
%at $T\sim 0.906$.
{}The $N=32$ data suggest
that all the observables depend continuously on $T$,
%both quantities are
%continuous at the critical temperature $T_{\rm c}$,
but their first derivatives with respect to $T$
seem to be discontinuous at $T \sim T_{{\rm c}1}$.
Thus we conclude that there exists a
second order phase transition at 
$T \sim T_{{\rm c}1}$.
%the phase transition between 
%may our Monte Carlo results suggest that the 
%phase transition is of second order.
%
%% \DOUBLEFIGURE[bht]
%% {E2.eps, width=.49\textwidth}
%% {R22.eps, width=.49\textwidth}
%% {Similar plot as fig.\ \ref{Fig_PL2}
%% for the internal energy $E$.
%% \label{fig:BHTE2}}
%% {Similar plot as fig.\ \ref{Fig_PL2}
%% for the ``extent of space'' $\langle R^2 \rangle$.
%% \label{Fig_R22}} 
  \FIGURE{
\epsfig{file=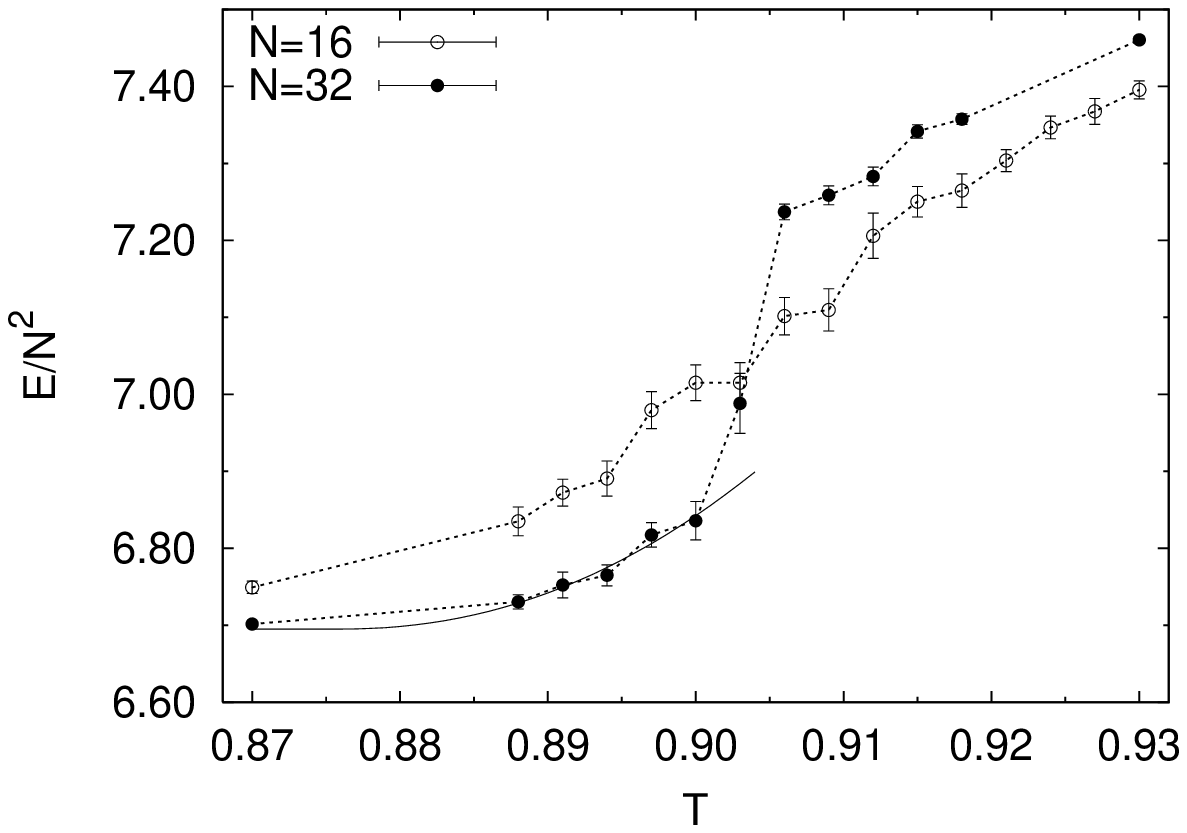,width=7.4cm}
\epsfig{file=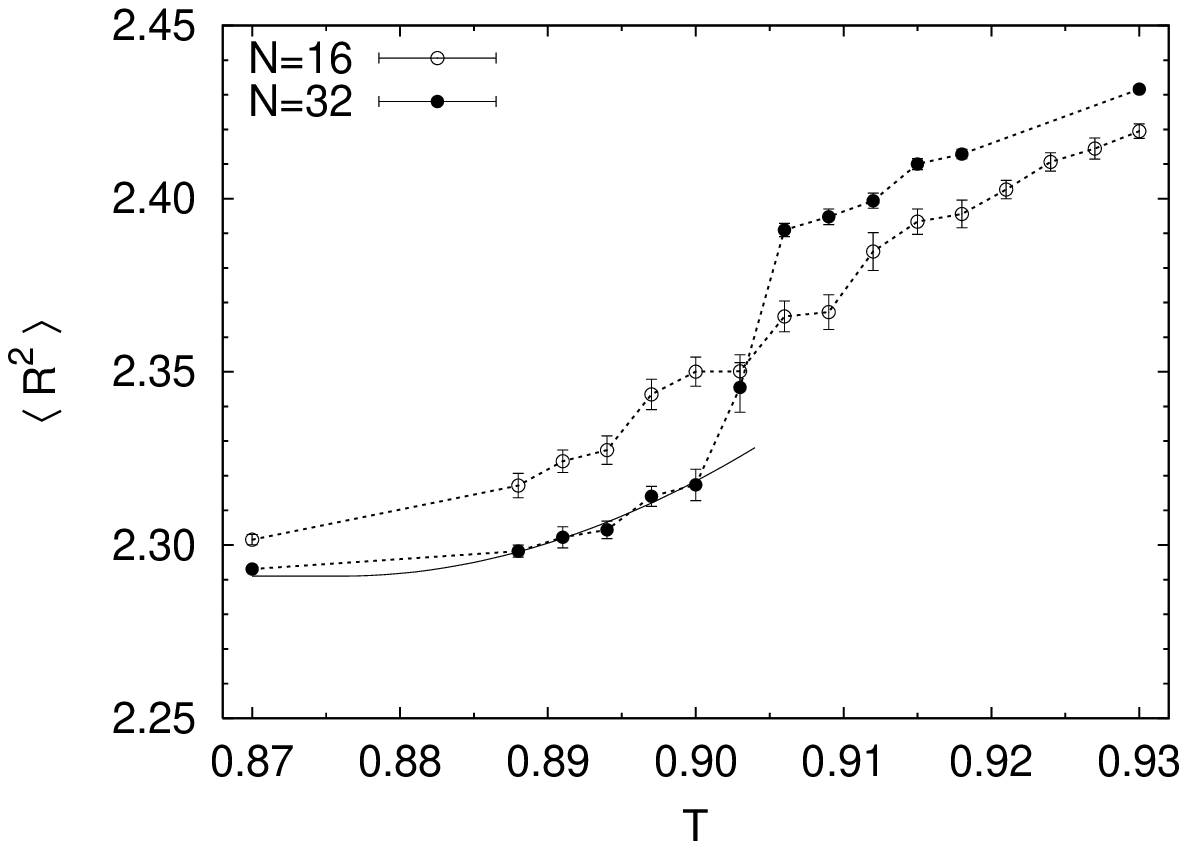,width=7.4cm}
   \caption{
The same as fig.\ \ref{fig:PL2} but
for the internal energy $E$ (left)
and for the ``extent of space'' $\langle R^2 \rangle$
(right).
The solid line in the left panel
is a fit to $\frac{E}{N^2} - \varepsilon_0 
= c (T - T_{{\rm c}2})^p$,
where $\varepsilon_0$ is given by the low-temperature data.
An analogous fit in the right panel yields
consistent values for $T_{{\rm c}2}$ and $p$.
%% with three free parameters
%% $T_{{\rm c}2}=0.878448$, $p=2.13294$ and 
%% $\alpha=413.216$. 
%% The constant $\varepsilon_0 = 6.695$
%% is the ground state energy (normalized by $N^2$)
%% obtained from the Monte Carlo data for $N=32$
%% at low temperature. 
%$T=0.900$, $T=0.903$, $T=0.906$ for $N=32$.
}
  \label{fig:BHTE2}
}

%% The Polyakov line (\ref{pol-def})
%% can be written in terms of the eigenvalues as
%% \beq
%% P = \frac{1}{N}
%% %\left\langle 
%% %\left| 
%% \sum_{j=1}^N \ee^{i \vartheta_{j}} \ .
%% %\right|  
%% %\right\rangle 
%% \label{orderparameter2}
%% \eeq
  \FIGURE{
\epsfig{file=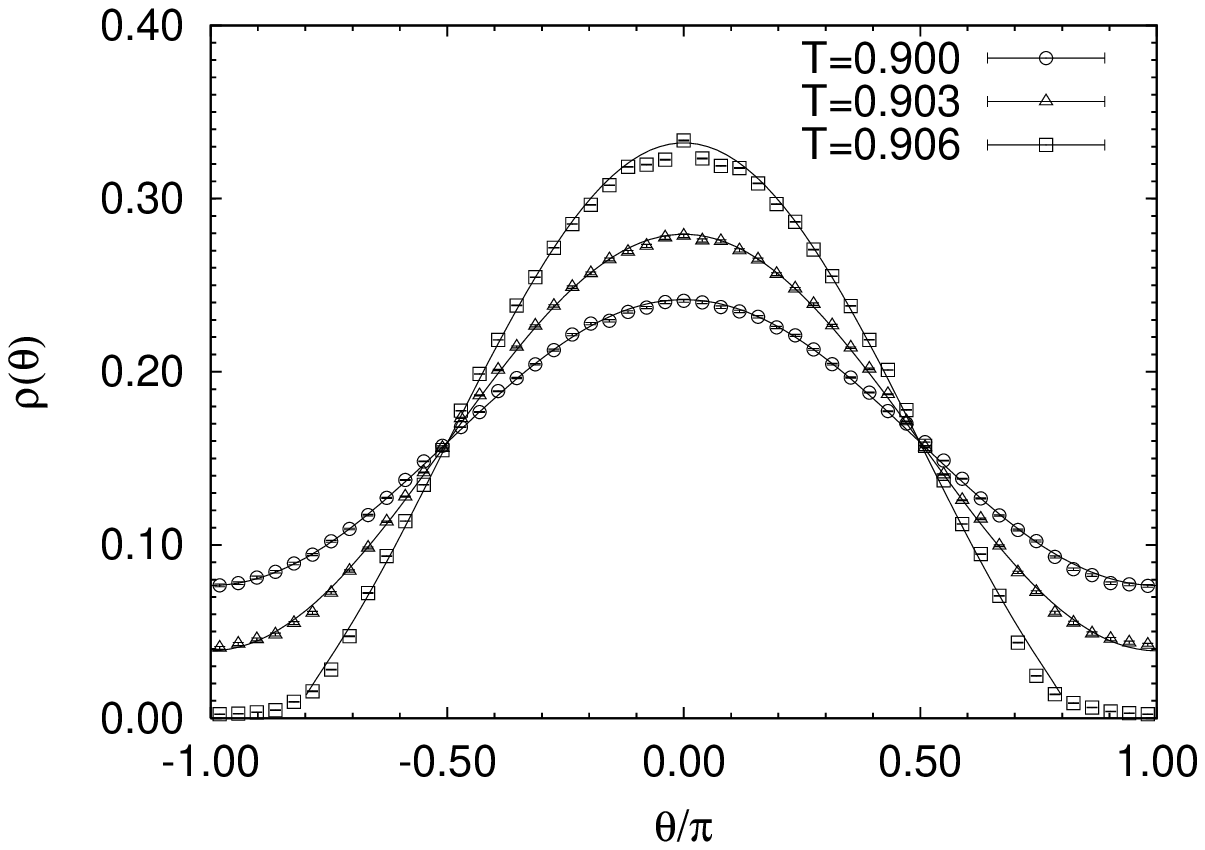,width=7.4cm}
\epsfig{file=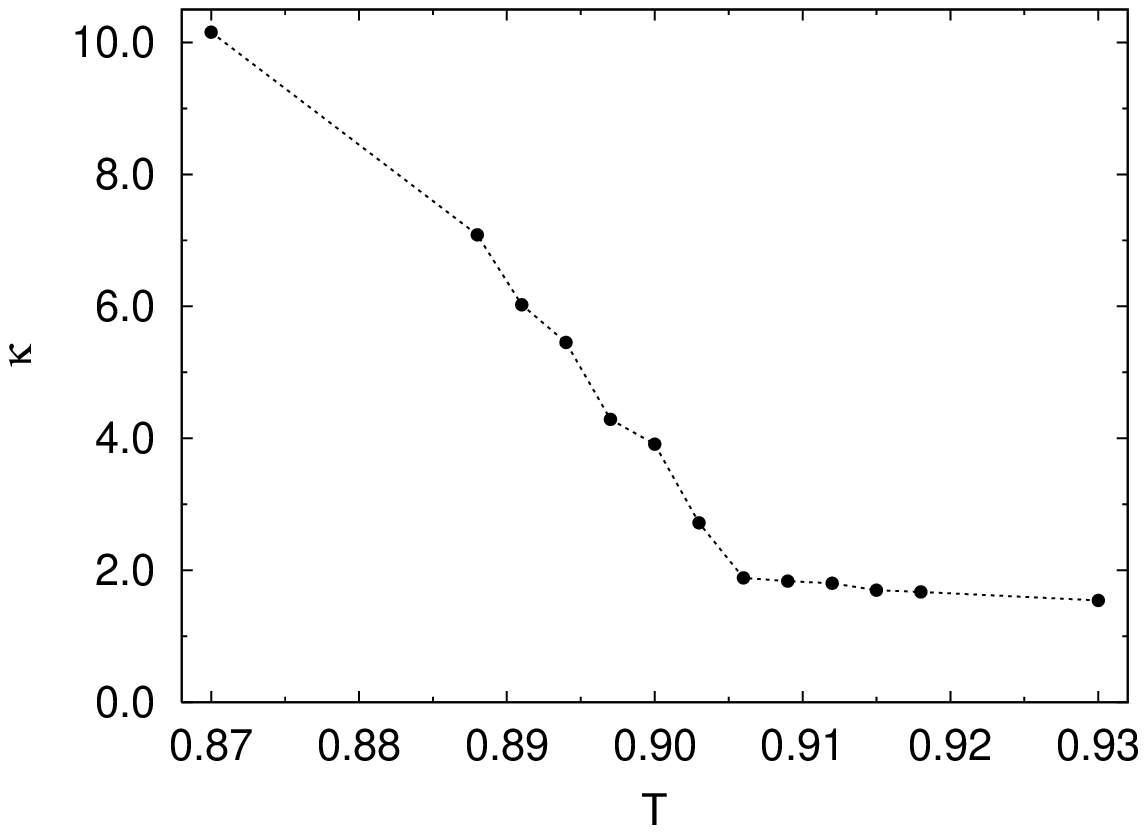,width=7.4cm}
%% lambda.eps, width=.49\textwidth
%% %7.4cm
%% }
%% \caption{
   \caption{(Left) The eigenvalue distribution 
$\rho(\theta)$ is plotted
for $N=32$ at
three different values of $T$ in the critical
regime. The solid lines represent a fit
to the Gross-Witten form given by either
(\ref{eigenvalue-distribution_exact})
or (\ref{eigenvalue-distribution_exact2})
depending on the fitting parameter $\kappa$.
%is chosen to optimize the fit
%at each $T$.
(Right) The value of $\kappa$ 
obtained by fitting 
the eigenvalue distribution
to the Gross-Witten form is
plotted against $T$.
The dotted line is drawn to guide the eye.
%$T=0.900$, $T=0.903$, $T=0.906$ for $N=32$.
}
\label{fig:EVD}
}

In order to clarify the nature of the phase
transition seen above, let us consider the
eigenvalues of 
the holonomy matrix $U$ given by 
eq.\ (\ref{pol-def2}),
which we denote as 
$e ^{i \vartheta_{j}}$ ($j = 1, \cdots , N$),
where 
$ \vartheta_{j} \in ( - \pi , \pi]$.
%$- \pi < \vartheta_{j} \le \pi$.
Then we define the eigenvalue distribution by
\beq
\rho (\theta) \equiv \frac{1}{N } 
\sum _{j=1}^N
\left\langle  \delta \left(\theta - 
\vartheta_{j} \right) \right\rangle \ ,
\label{def-rho}
\eeq
where we assume that the U(1) transformation (\ref{U1-sym})
is applied
%(\ref{U1-transf})
to each configuration in such a way that 
$\tr U$ becomes real positive
before taking the expectation value.\footnote{Note that
the U(1) transformation (\ref{U1-sym}) rotates
all the eigenvalues $\ee ^{i \vartheta_{j}}$ by some
constant angle on the complex plane.
If we took the expectation value in
eq.\ (\ref{def-rho}) naively,
we would trivially get a uniform distribution.
%constant $\rho(\theta)$ 
%due to the symmetry.
}
Using this definition, we have
\beq
\langle |P| \rangle =
\int_{-\pi}^{\pi}   d \theta \, 
\rho(\theta) \, \ee^{i\theta} \ .
\label{Pol-rho}
\eeq 
Fig.\ \ref{fig:EVD} (left) shows the result of
the eigenvalue distribution $\rho(\theta)$
for $N=32$.
It is clear that the second order phase transition 
is associated with the emergence
%development 
of a gap
in the eigenvalue distribution.
%% Since the U(1) transformation (\ref{U1-sym})
%% rotates all the eigenvalues
%% by a constant angle,
%% %the eigenvalue distribution should be uniform 
%% %if the U(1) symmetry is not spontaneously broken.
%% %
%% %% This is how the open Wilson line
%% %% serves as an order parameter.
%% %% In the broken phase, we have seen that
%% %% the open Wilson line $P$
%% %% acquires a non-zero expectation value.
%% the distribution calculated na
%% (\ref{def-rho})

Thus we find that the ``deconfined phase'',
in which the U(1) symmetry 
(\ref{U1-sym}) is spontaneously broken,
is further divided into the gapped phase
and the non-uniform phase.
%In the phases in which the U(1) symmetry 
%(\ref{U1-sym}) is spontaneously broken,
%
%let us consider
%Since the center U(1) symmetry is broken spontaneously
%on both sides of the phase transition, 
In these phases, we may consider
the Gross-Witten model \cite{GWW}
\beq
Z_{\rm GW}
= \int d U
\exp \left\{ \frac{N}{\kappa}
(\tr U + \tr U^\dag ) \right\}
\label{GWW-def}
\eeq
as a 
%simplified effective theory for the holonomy matrix, 
phenomenological model for
the holonomy matrix,
where the parameter $\kappa$ should be determined
as a function of $T$.
% near $T_{{\rm c}1}$.
The large-$N$ limit of
the eigenvalue distribution for the 
Gross-Witten model, which we denote as
$\rho_{\rm GW}(\theta)$, is known analytically \cite{GWW}.
%given explicitly.
% as follows.
For $\kappa < 2 $ we have a gapped distribution given by
\beq
\rho_{\rm GW} (\theta) =   
\frac{2}{\pi\kappa} \left(\cos\frac{\theta}{2} \right)
	\sqrt{\frac{\kappa}{2}-\sin^2\frac{\theta}{2}}
\label{eigenvalue-distribution_exact}
\eeq
for $| \theta | \le  2\sin^{-1} \sqrt{\frac{\kappa}{2}}$
and 0 otherwise.
For $\kappa \ge 2 $ we have a gapless distribution
\beq
\rho_{\rm GW}(\theta) = 
\frac{1}{2\pi} \left(
1 + \frac{2}{\kappa} \cos \theta
 \right) \ .
\label{eigenvalue-distribution_exact2}
\eeq
%The value of $\kappa$ obtained from the fit
%is plotted in fig.\ \ref{lambda-T}.
%
%For $N=16$ we observe that $\kappa$ crosses
%the critical value $2$ smoothly.
From these results, one obtains \cite{GWW}
\beq
\lim_{N \rightarrow \infty}
\left\langle \frac{1}{N} \tr U \right\rangle_{\rm GW}
= \int_{-\pi}^{\pi}   d \theta \, 
\rho_{\rm GW}(\theta) \, \ee^{i\theta} 
%|P| 
=   \left\{
\begin{array}{ll}
1 - \frac{\kappa}{4}
 & \mbox{~for~$\kappa<2$}  \  , \\
\frac{1}{\kappa}
&  \mbox{~for~$\kappa \ge 2$} \ , \\
\end{array}
\right. 
\label{pol}
\eeq
which crosses 1/2 at the critical point $\kappa=2$.
%The value of $\kappa$ obtained from the fit
%is plotted in fig.\ \ref{lambda-T}.
Note that (\ref{pol}) and its first derivative
with respect to $\kappa$ is continuous at 
$\kappa=2$, but the second derivative has a
discontinuity. Thus, the Gross-Witten
model undergoes a third order phase transition 
at $\kappa=2$.

As we see from
fig.\ \ref{fig:EVD} (left),
%We find that
the distribution $\rho(\theta)$ for $N=32$
can be nicely fitted to the Gross-Witten form
by choosing $\kappa$ appropriately at each $T$.
%as shown in fig.\ \ref{fig:EVD} (left).
The value of $\kappa$ obtained in this way
is plotted against $T$ on the right panel
of the same figure.
We observe that the first derivative of
$\kappa$ with respect to $T$ is discontinuous
at $T=T_{{\rm c}1}$.
This is reflected in the behavior 
of the Polyakov line 
$\langle |P| \rangle$ for $N=32$
shown in fig.\ \ref{fig:PL2}.
%changes the slope 
%at $T \sim 0.906$,
%as $T$ crosses $T_{{\rm c}1}$.
%when it crosses 1/2.
%Therefore, it is not a contradiction to 
Thus it is possible to
obtain a second (instead of third) order phase transition
between the non-uniform phase and the gapped phase
in the present model, despite the fact that the 
eigenvalue distribution is well described 
by the Gross-Witten form.

%% \beq
%% \rho(\theta) =   \left\{
%% \begin{array}{cl}
%% \frac{2}{\pi\kappa} \cos\frac{\theta}{2} 
%% 	\sqrt{\frac{\kappa}{2}-\sin^2\frac{\theta}{2}}
%% %	\sqrt{1-\sqrt{1-\frac{1}{f_1(\beta)}}-\sin^2\frac{\theta}{2}} 
%% & \mbox{~for~$| \theta | \le \theta_{0} $}  \\
%% 0 & \mbox{~for~$  |\theta| > \theta_{0} $}  \\
%% \end{array}
%% \right.
%% \label{eigenvalue-distribution_exact}
%% \eeq
%
%Namely, we can fit the distribution by 
%the above form by choosing $\kappa$
%at each temperature in the critical regime, 
%
%We take this as an evidence for the
%large-$N$ behavior.
%
%% \FIGURE{
%% \epsfig{file=
%% lambda.eps, width=.49\textwidth
%% %7.4cm
%% }
%% \caption{
%% The value of $\kappa$ 
%% obtained by fitting the eigenvalue distribution
%% to the Gross-Witten form is
%% plotted against temperature. 
%% The solid lines are drawn to guide the eye.
%% }
%% \label{lambda-T}
%% }
%% On the other hand, the Polyakov line 
%% $\langle |P| \rangle$ for $N=32$
%% shown in fig.\ \ref{fig:PL2}
%% changes the slope 
%% %at $T \sim 0.906$,
%% as $T$ crosses $T_{{\rm c}1}$.
%% %when it crosses 1/2.
%% In fact the value of $\kappa$ obtained
%% by fitting $\rho(\theta)$
%% %the Polyakov line 
%% to the Gross-Witten form
%% has analogous dependence on $T$.
%% Therefore, it is not a contradiction to 
%% obtain a second (instead of third) order phase transition
%% between the non-uniform phase and the gapped phase
%% in the present model.

In the ``confined phase'', the U(1) symmetry 
(\ref{U1-sym}) is unbroken, and therefore
the eigenvalue distribution 
$\rho(\theta)$ 
is uniform in the large-$N$ limit;
hence we call it the uniform phase following
the present terminology.
However, 
as we have seen in fig.\ \ref{Fig_PL},
the Polyakov line, which should vanish
for a uniform distribution,
%is expected to vanish in the large-$N$ limit, 
seems to be of O($1/N$) at low temperature,
which actually vanishes slowly
with increasing $N$.
% in the large-$N$ limit.
%This makes it difficult to identify the uniform phase
%directly from the eigenvalue distribution or the
%Polyakov line due to finite $N$ effects.
%
%In particular, it seems almost impossible
%to distinguish
%the uniform phase and the non-uniform phase
%from the Polyakov line.
On the other hand, 
the Eguchi-Kawai equivalence, which is
a consequence of the unbroken U(1) symmetry,
%in the large $N$ limit,
holds with high accuracy
%is clearly observed 
at low temperature as we have seen in fig.\ \ref{fig:BHTE}.
In general the breaking of 
the Eguchi-Kawai equivalence due to finite $N$ effects
is expected to be of O($1/N^2$).
%, which naturally explains our observation.
%This implies that 
%the Eguchi-Kawai equivalence 
%is less affected by
%finite $N$ effects than
%the eigenvalue distribution.
Taking advantage of this fact,
%observation,
we identify the uniform phase
using the Eguchi-Kawai equivalence
instead of identifying it directly using
the eigenvalue distribution.
%As we have seen in fig.\ \ref{fig:BHTE},
% shows that
In particular, the internal energy normalized by $N^2$
% and the extent of space are 
is constant at low temperature,
and the constant value,
which gives the ground state energy,
is extracted as
$ \varepsilon_0 = 6.695(5)$ from the $N=32$ data.
With increasing $T$, one enters the non-uniform
phase at some critical point $T_{{\rm c}2}$,
where the internal energy starts to deviate from 
%$ \varepsilon_0$.
this constant value.
We fit the results of the internal energy 
for $N=32$ shown in fig.\ \ref{fig:BHTE2}
to the behavior
\beq
\frac{E}{N^2}  - \varepsilon_0 = 
   \left\{
\begin{array}{ll}
0  & \mbox{~for~$T \le T_{{\rm c}2}$}  \  , \\
%\varepsilon_0 + 
c (T - T_{{\rm c}2})^p
&  \mbox{~for~$T > T_{{\rm c}2}$} \ , \\
\end{array}
\right. 
\label{energy-3rd}
\eeq
%\varepsilon_0 + C (T - T_{{\rm c}1})^p
with three free parameters, which are determined as
%$c=4.0(1)\times 10^2$,
%$T_{{\rm c}2}=0.8760(3)$ and $p=2.12(4)$.
%$c=4(3)\times 10^2$,
$c=413\pm 310$,
$T_{{\rm c}2}=0.8758(9)$ and $p=2.1(2)$.
We redo similar analysis for the ``extent of space''
$\langle R^2 \rangle$.
The constant value at low temperature is given by
$ (r_0)^2 = 2.291(1)$.
The deviation $\langle R^2 \rangle 
- (r_0)^2 $ can be fitted
to the behavior (\ref{energy-3rd})
with three free parameters, which are determined 
this time as
%$c=4(3)\times 10$,
$c=39 \pm 36$,
$T_{{\rm c}2}=0.8763(4)$ and $p=1.9(2)$.
%$c=39(1)$,
%$T_{{\rm c}2}=0.8762(2)$ and $p=1.942(9)$.
The results for $T_{{\rm c}2}$ and $p$ obtained
from the two observables are consistent
with each other.
%assuming that the fitting error for $p$ is
%slightly underestimated.
Moreover they suggest $p=2$, which implies that
the phase transition
%critical point
between the uniform phase and the non-uniform phase
is 
%$T_{{\rm c}2}=0.878$ and the phase transition is 
of third order.

Thus we have identified 
the non-uniform phase between the two critical points
$T_{{\rm c}1} = 0.905(2)$ and $T_{{\rm c}2}=0.8761(3)$.
The range of $T$ is very narrow.
This is not so surprising, however,
%However, we do not consider this as a mistery,
given that the Polyakov line grows very rapidly
as we increase the temperature in the critical regime.
Note that the Polyakov line
should lie within the range $[0,\frac{1}{2}]$
in the non-uniform phase according to the 
(successful) phenomenological description 
in terms of the Gross-Witten model.

Let us also comment that the phase structure 
obtained above
is consistent with one of the two scenarios suggested
by the Landau-Ginzburg analysis \cite{Aharony:2004ig}.
The other scenario was that
the Polyakov line jumps from 0 to 1/2
at some critical temperature, indicating
a single first order transition between the uniform
phase and the gapped phase.
This behavior was observed in the plane-wave
matrix model at finite temperature \cite{FSS,KNY}.
On the other hand, in ref.\ \cite{Kawahara:2007nw}
we observed no phase transition in the fuzzy
sphere background.
%phase. 
This might also be the case
with the supersymmetric version 
of the present model \cite{Barbon:1998cr,Aharony4}.

%% %\FIGURE[bht]
%% \DOUBLEFIGURE[bht]
%% {Eigen_N16.eps, width=.49\textwidth}
%% {Eigen_N32.eps, width=.49\textwidth}
%% %lambda.eps
%% {Eigenvalue distribution plotted for
%% $T=0.900$, $T=0.903$, $T=0.906$ for $N=32$.
%% \label{fig:EVD}}
%% {The value of $\lambda$ 
%% obtained by fitting the eigenvalue distribution
%% to the Gross-Witten form is
%% plotted against temperature. 
%% The solid lines are drawn to guide the eye.
%% \label{lambda-T}} 

%%%

%% The Polyakov line can be obtained as a function of
%% $\lambda$ as
%% \beq
%% \rho(\theta) =   \left\{
%% \begin{array}{cl}
%% \frac{1}{\lambda}
%% &  \mbox{~for~$\lambda>2$} \ , \\
%% 1 - \frac{\lambda}{4}
%%  & \mbox{~for~$\lambda<2$}  \, , \\
%% \end{array}
%% \right. \ .
%% \label{pol}
%% \eeq

%%%%%%%%%%%%%%%%%%%%%%%%%%%%%%%%%%

\section{Connection to the black-hole/black-string transition}
\label{section:GL}

In fact one can show that the model (\ref{action})
describes the high temperature regime
of (1+1)d super Yang-Mills theory on a circle.
At low-temperature strong-coupling regime,
the 2d theory is expected to have
a first order phase transition,
which corresponds to
the black-hole/black-string transition
in the dual gravity description.
In this section we review this connection,
and discuss the implication of our results on it.

  \FIGURE{
\epsfig{file=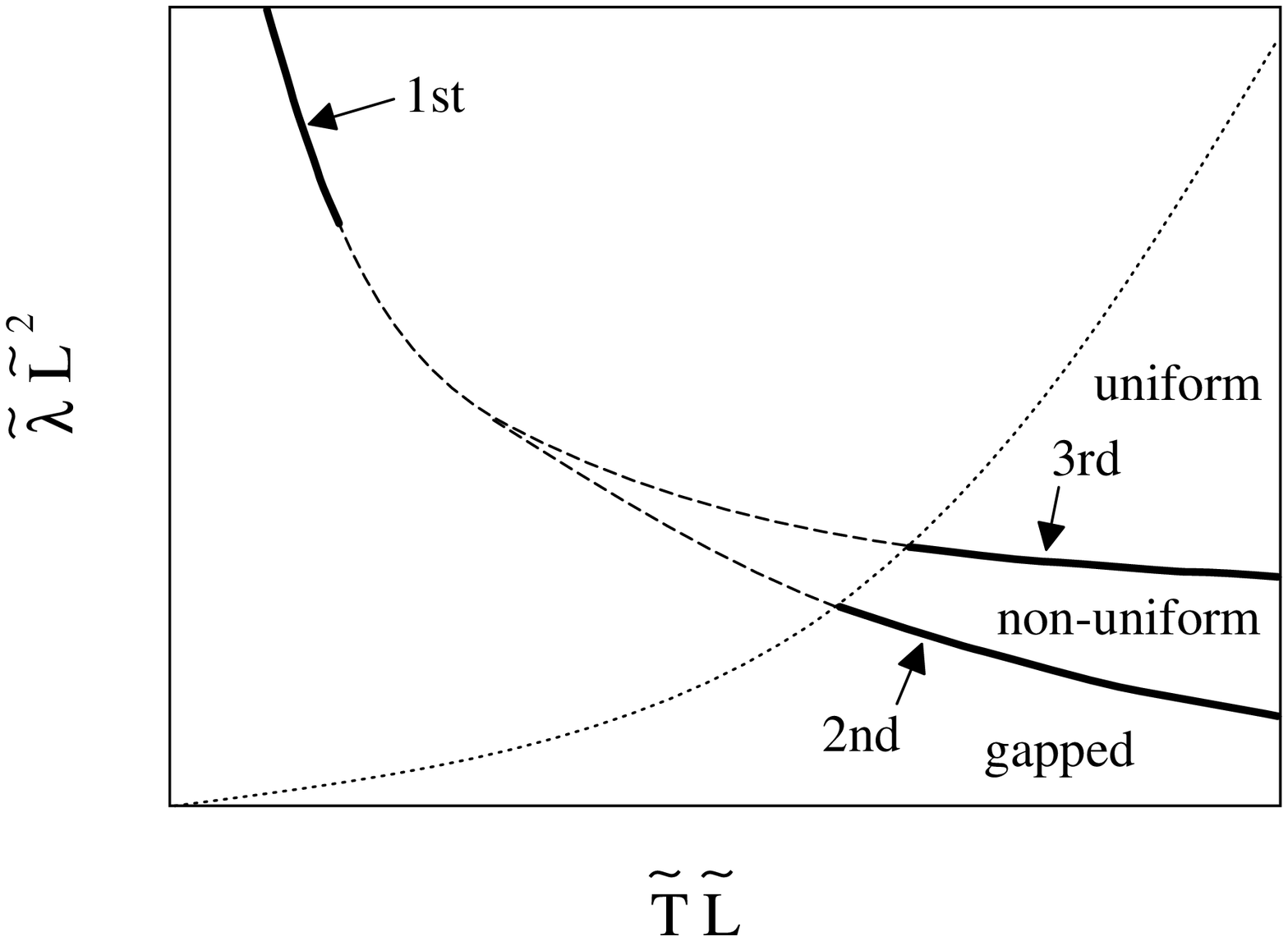,width=.8\textwidth}
%% lambda.eps, width=.49\textwidth
%% %7.4cm
%% }
%% \caption{
   \caption{
A schematic view of the phase diagram
of the 2d super Yang-Mills theory
in terms of dimensionless parameters
$\tilde{T}\tilde{L}$ and 
$\tilde{\lambda}\tilde{L}^2$.
The region
below the dotted line  (\ref{decoupledKK})
can be
well described by our 1d model, from which
we obtain the critical lines 
corresponding to (\ref{critical-points}).
The upper left corner is conjectured to
have a dual gravity description, which predicts
the first order phase transition at (\ref{grav-1st}).
The dashed lines represent our speculation
that the non-uniform phase ceases to exist
at a tri-critical point for consistency
with the prediction from the gauge/gravity duality.
}
\label{fig:PfSt}
}

%In ref.\cite{Aharony:2004ig}
%the same model has been studied 
Let us consider 2d U($N$) ${\cal N}=8$ super 
Yang-Mills theory,\footnote{We put tildes on all the parameters
of the 2d theory to distinguish them from the parameters
of our 1d theory.}
which can be obtained by dimensionally reducing
10d U($N$) ${\cal N}=1$ super Yang-Mills theory 
to 2d.
In order to put the theory at finite temperature,
we compactify the Euclidean time direction
to a circle with the circumference 
$\tilde{\beta}$, which corresponds to the inverse temperature
$\tilde{\beta}= \frac{1}{\tilde{T}}$.
Furthermore we compactify the spatial direction to 
a circle\footnote{This model is formally 
the same as the matrix string theory
\cite{Motl:1997th,Banks:1996my,DVV}
except that fermions obey anti-periodic boundary conditions
in the temporal direction.}
with the circumference $\tilde{L}$.
%Since we have 3 dimensionful parameters $\tilde{\beta}$,
%$\tilde{L}$ and $\tilde{\lambda}$, 
At sufficiently high temperature,
the temporal Kaluza-Klein modes decouple,
and one obtains our model (\ref{action})
with appropriate identification of parameters. 
%and the dimensional reduction takes place.
Note in particular that the fermions in the 2d theory
decouple due to anti-periodic boundary conditions in the temporal
direction, and therefore one ends up with a bosonic theory.
The condition for the decoupling of 
the temporal Kaluza-Klein modes 
is given by 
%$\tilde{T} \gg \sqrt[3]{\frac{\tilde{\lambda}}{\tilde{L}}}$.
\beq
\tilde{T} \gg 
\left(\frac{\tilde{\lambda}}{\tilde{L}}\right)^{1/3} \ .
\label{decoupledKK}
\eeq

In this interpretation
the $t$-direction of our 1d model
is identified with the spatial direction
of the original 2d theory, and the relationship 
among the parameters is given by
\beq
\beta = \tilde{L} \ , \quad
\lambda = \tilde{\lambda} \tilde{T} \ .
%\ .
\eeq
Hence the dimensionless effective coupling constant
$\lambda_{\rm eff}$ defined in eq.\ (\ref{lam-eff}) is written as
\beq
\lambda_{\rm eff} =  \tilde{\lambda} \tilde{T} \tilde{L}^3  \ .
\eeq
Note also that the Polyakov line (\ref{pol-def}) 
should be regarded
as the Wilson loop winding around a spatial direction
in the 2d theory.
%% Therefore the dimensionless coupling constant 
%% (\ref{lam-eff}) can be written in terms of the parameters
%% of the 2d theory as
%% $\lambda_{\rm eff} =  \tilde{\lambda} \tilde{T} \tilde{L}^3 $.
In ref.\ \cite{Aharony:2004ig} the 1d model (\ref{action})
was studied from this point of view and a phase transition
was observed around $\lambda_{\rm eff} \simeq 1.4$.
Our results suggest that actually there exist two
phase transitions at
\beq
\lambda_{\rm eff}
%=\lambda_{\rm eff}^{({\rm c}1)} 
\simeq 1.35(1) \ 
\mbox{~and~}
%, \quad
\lambda_{\rm eff}
%\lambda_{\rm eff}^{({\rm c}2)} 
\simeq 1.487(2) \ ,
\label{critical-points}
\eeq
which are of second order and
of third order, respectively.
%% a second order phase transition
%% at $\lambda_{\rm eff} \simeq 1.35(1)$,
%% and a third order phase transition
%% at $\lambda_{\rm eff} \simeq 1.487(1)$.

At large $\tilde{\lambda}$
and small $\tilde{T}$,
the 2d theory has a dual gravity description,
which predicts
a first order phase transition at \cite{Aharony:2004ig}
\beq
\tilde{T}\tilde{L}
= \frac{2.29}{\sqrt{\tilde{\lambda} \tilde{L}^2}} \ .
\label{grav-1st}
\eeq
On the gravity side, this 
corresponds to
the black-hole/black-string transition,
which is associated with 
the Gregory-Laflamme instability \cite{Gregory:1993vy}
of the black string winding around the spatial direction.
The black-string phase and the black-hole phase
can be naturally identified with the
uniform phase and the gapped phase on the gauge
theory side.
A similar correspondence was suggested 
earlier by ref.\ \cite{Susskind:1997dr}.
%As temperature rises the transition (similar to
%our transition) might occurs. 
%
%It is therefore interesting that we are able
%to identify the three phases in the present model.

%Our conclusion suggests 
We speculate that 
the first order phase transition
predicted at low temperature
actually splits into two continuous transitions
as one increases the temperature.
In figure \ref{fig:PfSt} we present
a schematic view of the phase diagram
of the 2d super Yang-Mills theory
that emerges from the present work.
%the two transitions
%in the high $T$ limit of the 2d theory
%actually merges into one as one lowers the temperature.

\section{Summary and discussions}
\label{Summary}

In this paper we have investigated the
phase structure
of matrix quantum mechanics at finite temperature.
We have identified three phases.
At high temperature,
%In the gapped phase, 
the high temperature expansion up to the next-leading
order gives a precise description of various 
observables.
At low temperature,
%In the uniform phase, 
the internal energy,
in particular, does not depend on the temperature 
as a consequence of the Eguchi-Kawai equivalence.
This property enables us to determine the
critical point and the order of the
transition between the uniform phase and
the non-uniform phase.
%% We have determined the ground state energy accurately.
%% we can discuss the gauge/gravity correspondence
%% by interpreting the same model as the high temperature
%% limit of the 2d ${\cal N}=8$ super Yang-Mills theory.
%% We hope that our results shed light on the 
%% phase diagram of the super Yang-Mills theory,
%% %which is interesting on its own right,
%% and possibly also on the physics of black objects
%% in the dual gravity theory.
%
%An analogous phase has been found earlier 
%as the end-point of the Gregory-Laflamme 
%instability \cite{Gregory:1993vy}.
%That it has a natural
%realization in the present model
%is interesting from the view point of the gauge/gravity 
%correspondence.
%
%The phase transition between the
%non-uniform phase and the localized phase
%appears to be of second order.
%
%Based on 

%The quality of the observed
%Eguchi-Kawai equivalence at low temperature, and 
In the non-uniform phase and the gapped phase,
the eigenvalue distribution of the
holonomy matrix can be fitted nicely
to the Gross-Witten form.
While this suggests that we are 
already seeing the large-$N$ behaviors,
%The reasonable fits of $\rho(\theta)$
%to the Gross-Witten form suggest that we are 
%seeing the large-$N$ physics.
%However, 
it is certainly desirable to
confirm the stability of our results against
increasing $N$,
which we leave for future investigations.
%in order to draw more definite 
%conclusions, we need to study larger $N$ and
%confirm the stability of our results.
%This is beyond the scope of the present paper,
%and we leave it for future investigations.

Our results can be alternatively interpreted
as representing the high temperature behavior
of 2d ${\cal N}=8$ super Yang-Mills theory.
%Viewing the present model as the high temperature
%limit of the 
%2d ${\cal N}=8$ super Yang-Mills theory,
%our results
% shed light on the 
%have implications on the 
%phase diagram of the super Yang-Mills theory.
%which is interesting on its own right,
%and possibly on the physics of black objects
%in the dual gravity theory.
The low temperature behavior of 
%the same theory 
that theory
can be predicted by
the gauge/gravity correspondence.
% and the non-uniform phase does not appear there.
%
% of the same 2d theory.
%is not expected to appear.
%The low temperature regime of
%the same theory 
%
%the phase diagram,
%on the other hand, 
%can be studied by the gravity
%theory through the gauge/gravity correspondence.
%The parameter region studied by the present work 
%and that by the gauge/gravity correspondence
%lie at the opposite ends of the phase diagram.
%There, the non-uniform phase is not expected to appear.
%Therefore,
%
%If our conclusion represents the large-$N$ limit
%at least qualitatively, then it 
%our conclusion implies that 
%It is therefore suggested 
We speculate that the non-uniform phase 
identified in 
the present paper
%this paper
ceases to exist below
%appears at 
some temperature in the phase diagram
of the 2d
%$(1+1)$d 
super Yang-Mills theory.
If so,
it is interesting to locate this point explicitly.
%Apart from such interpretation,
%we may 

The non-uniform phase may exist also
in higher dimensional bosonic gauge theories
%pure Yang-Mills theory 
on a finite torus
\cite{Narayanan:2003fc,Kiskis:2003rd}.
That will have implications on 
the phase diagram
of super Yang-Mills theories in $D=3,4$.
%higher dimensions.
The low temperature regime of these theories
is discussed in ref.\ \cite{Hanada:2007wn} 
based on the gauge/gravity correspondence.

%{}From the viewpoint of the gauge/gravity correspondence,
%it would also be interesting
%to study the supersymmetric version of the present model
%using the non-lattice
%simulation proposed recently \cite{non-lattice}.
It would be also interesting
to study the supersymmetric version of the present model
using the non-lattice
simulation proposed recently \cite{non-lattice}.
In particular, by studying the low temperature regime,
we would be able to test the predictions of 
the gauge/gravity correspondence directly.
The phase transitions are expected to disappear
\cite{Barbon:1998cr,Aharony4},
and the internal energy is expected to vanish as 
$T \rightarrow 0$ with the power law behavior
\cite{KLL} obtained from the dual black-hole 
geometry \cite{Klebanov:1996un}.
%The uniform phase
%is expected to disappear 
%(or to be pushed towards $T=0$) 
%\cite{Barbon:1998cr,Aharony4}.
%The internal energy is expected to vanish as 
%$T \rightarrow 0$ following the power law behavior
%\cite{KLL} obtained for the dual black-hole 
%geometry \cite{Klebanov:1996un}.
Indeed the preliminary results 
for a simplified model with 4 supercharges
\cite{non-lattice} agree qualitatively with these
expectations.
Studies of the model with 16 supercharges
%along the same line 
are in progress \cite{Mtheory}.

\section*{Acknowledgments} 

We would like to thank
%Takehiro Azuma, 
Kazuyuki Furuuchi,
%Masanori Hanada,
Norihiro Iizuka,
%Shun'ya Mizoguchi, 
Satoshi Iso,
Gordon Semenoff
and Kentaroh Yoshida 
for useful comments and discussions.
We are also grateful to Niels Obers
and Troels Harmark for drawing our
attention to ref.\ \cite{Harmark:2002tr},
which was overlooked in the earlier version
of this paper.

\appendix

\section{Derivation of a formula for the internal energy}
\label{section:EO}

In this section we derive the formula (\ref{E=dbF})
relating the internal energy of the present model
to the expectation value (\ref{F2def}),
which is directly accessible by Monte Carlo simulation.
%which is related to the first derivative of 
%the free energy with respect to the temperature.

Let us first rewrite (\ref{defE}) as
\beq
E = - \frac{1}{Z(\beta)}
\lim_{\Delta \beta \rightarrow 0 } 
\frac{Z(\beta ') - Z(\beta)}{\Delta \beta} \ ,
\label{E-cal}
\eeq
where $\beta ' = \beta + \Delta \beta$,
and represent $Z(\beta ')$ for later convenience as
\beq
Z(\beta ') = \int [{\cal D} X ']_{\beta '}
[{\cal D} A']_{\beta '} \, \ee^{- S'}  \ ,
\eeq
where $S'$ is obtained from $S$ given in (\ref{action})
by replacing $\beta$, $t$, $A(t)$, $X_i(t)$
with $\beta '$, $t'$, $A'(t')$, $X_i '(t')$.
In order to relate $Z(\beta ')$ to $Z(\beta)$,
we consider the transformation
\beq
t ' = \frac{\beta '}{\beta} \, t \ ,
\quad
A '(t ') = \frac{\beta}{\beta '} \, A(t) \ ,
\quad
X_i ' (t ') = \sqrt{\frac{\beta '}{\beta}} \, X_i (t) \ .
\eeq
The factors in front of the fields are
motivated on dimensional grounds, and in particular
we have
$ [ {\cal D} X ' ]_{\beta ' } = [ {\cal D} X]_{\beta}$
and
$[ {\cal D} A ' ]_{\beta ' } = [ {\cal D} A]_{\beta} $.
Under this transformation, 
the kinetic term in $S'$ reduces to that in $S$,
but the interaction term transforms non-trivially as
%% Note also that the kinetic term in the action
%% (\ref{action}) is invariant under this transformation.
%% The interaction term is not invariant, though,
%% and we obtain
\beq
\int_0^{\beta '} \!\!dt ' \, \tr
%\frac{1}{4}
\Bigl( [X_i ' (t ' ),X_j ' (t ' )]^2  \Bigr)
= \left(\frac{\beta ' }{\beta} \right)^3
\int_0^\beta  \!\!dt  \, \tr
%\frac{1}{4}
\Bigl( [X_i  (t  ),X_j (t )]^2 \Bigr) \ .
\eeq
This gives us the relation
% between $Z[\beta']$ and $Z[\beta]$
\beqa
\label{Z/Z}
Z(\beta')=Z(\beta) \left\{
1-\frac{3}{4}  N^2 \Delta \beta  \langle F^2 \rangle
%\left\langle F^2 \right\rangle 
+{\rm O} \Bigl( (\Delta \beta)^2 \Bigr)
\right\}  \  ,
\eeqa 
where the operator $F^2$ is defined by (\ref{F2def}).
Plugging this into (\ref{E-cal}), we get
(\ref{E=dbF}).

\section{Details of Monte Carlo simulation}
\label{section:algorithm}

In this section
we present the details of
%explain the algorithm used for
our Monte Carlo simulation.
We discretize the Euclidean time direction 
and obtain the partition function
\beqa
\label{LAZ}
Z_{\rm lat} &=& \int [dV][dX_i]\exp(-S_{\rm lat}) \\
\label{LAA}
S_{\rm lat} &=& aN \sum_{n=1}^{N_t}\tr
\Bigg\{\frac{1}{2} 
\left(
\frac{V(n)X_i(n+1)V^{\dagger}(n)-X_i(n)}{a}
\right)^2
 -\frac{1}{4}
%\Big(
[X_i(n),X_j(n)]^2 
%\Big)
\Bigg\} \ ,
\eeqa 
where $a$ is the lattice spacing, and 
the inverse temperature is given by 
$\beta=a N_t$.
The link variables $V(n)$ are $N \times N$ 
unitary matrices representing the gauge connection
between the lattice sites.
Due to the periodic boundary conditions,
we have $X_i(N_t+1)=X_i(1)$.

Although it is possible to simulate
the system (\ref{LAZ}) directly,
let us simplify it \cite{Aharony:2004ig} by taking the
static diagonal gauge 
\beq
V(1)=V(2)=\cdots =V(N_t) 
\equiv V
= {\rm diag} \Big( \ee^{i \theta_1 / N_t},
\ee^{i \theta_2 / N_t},\cdots,
\ee^{i \theta_N /N_t} \Big)\,.
\label{diagonalized A}
\eeq
The integration measure for the angular variables $\theta_a
\in (-\pi , \pi]$ is given by 
\beqa
[d \theta] &=& \left( 
\prod_{a=1}^N  d \theta_a \right) \, 
\Delta (\theta)  \ , \\
\Delta (\theta) &=& 
 \prod_{a < b} \, \sin ^ 2 \Big(\frac{\theta_a-\theta_b}{2}\Big) \ ,
\eeqa
where $\Delta (\theta)$ is the Vandermonde determinant.

The operators (\ref{pol-def2}),
(\ref{F2def}) and (\ref{def_R2})
can be calculated on the lattice by 
\beqa
%P  &=&
%\frac{1}{N} \sum_{a=1}^N \ee ^{i \theta_a} \ ,\\
U  &=& V(1) V(2) \cdots V(N_t) = 
{\rm diag} \Big( \ee^{i \theta_1},
\ee^{i \theta_2}, \cdots , \ee^{i \theta_N} \Big)  \  ,\\
%\frac{1}{N} \sum_{a=1}^N \ee ^{i \theta_a} \ ,\\
F^2
&=&
-\frac{1}{N N_t}\sum^{N_t}_{n=1}\tr\Big([X_i(n),X_j(n)]^2\Big) \ ,\\
R^2
&=&\frac{1}{N N_t}\sum^{N_t}_{n=1}\tr\Big(X_i(n)^2\Big) \ .
\eeqa

In order to employ the heat-bath algorithm 
for updating $X_i(n)$, we use the trick used in simulating
the bosonic IKKT model \cite{HNT}.
Namely, we introduce 
$N \times N$ hermitian matrices
$Q_{ij}(n)$
($i,j=1,\cdots , 9$; $i<j$) as
auxiliary fields,
% obeying the condition $Q_{ij}(n) = Q_{ji}(n)$.
and consider the partition function
\begin{eqnarray}
\tilde{Z}_{\rm lat} &=& \int [d \theta][dX][dQ]
\exp(-\tilde{S}_{\rm lat}) \\
\tilde{S}_{\rm lat} &=& aN \sum^{N_t}_{n=1} \tr \left[
\frac{1}{2}
\left(\frac{VX_i(n+1)V^{\dagger}-X_i(n)}{a}\right)^2 
\right. \nonumber \\
&& \left. +\frac{1}{2} \sum_{i<j} \Bigl(
Q_{ij}(n)^2-2Q_{ij}(n) \{ X_i(n),X_j(n) \}
\Bigr)  
+2\sum_{i<j}X_i(n)^2 X_j(n)^2 \right] \ .
\label{action-aux}
\end{eqnarray}
%where $G_{ij}(n) \equiv \{ X_i(n),X_j(n) \}$.
Integrating out the auxiliary fields $Q_{ij}(n)$,
we retrieve the original action (\ref{LAA}).
We update $Q_{ij}(n)$ and $X_i (n)$ for each $n$ 
in the same way as described in ref.\ \cite{HNT}.

After updating all the elements of 
$Q_{ij}(n)$ and $X_i (n)$ for each $n$,
we update the angular variables $\theta_a$ using
the standard Metropolis algorithm.
Typically the acceptance rate is not very high.
We therefore repeat the Metropolis procedure
sufficiently many times so that most of
$\theta_a$ get updated.
This defines our ``one sweep''.
For $N=16$ ($N=32$) we have made 200,000 sweeps 
(150,000 sweeps) in total,
and discarded the first 20,000 sweeps (40,000 sweeps)
for thermalization at each temperature.
The simulation has been performed on
PCs with Pentium 4 (3GHz), and it took a few weeks
to get results at each temperature for $N=32$.

%It is necessary to reduce the systematic error 
%due to the discretization in order to give 
%reveal the properties
%discussed in this section.
%
% is therefore
%expected to be largely reduced in the present work.
%
%This is crucial in unraveling some properties of
%this model and in quantitative comparison with results
%obtained by different methods.

%%%%%%%%%%%%%%%%%%%%%%%%%%%%%%%%%%%%%%%%%%%%%%%%%%%%%%%%%%%%%%%%%%%%%%

\end{document}